\documentclass[letterpaper,twocolumn,10pt]{article}
\usepackage{usenix}

\usepackage{graphicx} 
\usepackage{siunitx}
\usepackage{color}
\usepackage{booktabs}
\usepackage{tikz}
\usepackage{xspace}
\usepackage{caption}
\usepackage{subcaption}
\usepackage[capitalize]{cleveref}
\usepackage{tabularx}
\usepackage{authblk}

\NewDocumentCommand\system{}{\textsc{Stabl}\xspace}

\NewDocumentCommand\systemlong{}{Sensitivity Testing and Analaysis for BLockchain\xspace}

\NewDocumentCommand\diablo{}{{\sc Diablo}\xspace}

\NewDocumentCommand\score{}{sensitivity score\xspace}
\NewDocumentCommand\Score{}{Sensitivity score\xspace}

\NewDocumentCommand\resilience{}{resilience\xspace}
\NewDocumentCommand\Resilience{}{Resilience\xspace}

\NewDocumentCommand\recoverability{}{recoverability\xspace}
\NewDocumentCommand\Recoverability{}{Recoverability\xspace}

\NewDocumentCommand\Ptolerance{}{Partition Tolerance\xspace}

\NewDocumentCommand\btolerance{}{Byzantine fault tolerance\xspace}
\NewDocumentCommand\Btolerance{}{Byzantine Fault Tolerance\xspace}

\usepackage{marginnote}

\newrobustcmd{\tagbox}[2]{\colorbox{#1}{\bfseries\sffamily\footnotesize\textcolor{white}{#2}}}
\newrobustcmd{\annoteMargin}[3]{{\hspace{-.5mm}\marginnote{\tagbox{#3}{#2}}}}
\newrobustcmd{\annoteInline}[3]{{\tagbox{#3}{#2} \color{#3}{#1}}}
\newrobustcmd{\annote}[3]{\annoteMargin{}{#2}{#3}\annoteInline{#1}{#2}{#3}}

\title{\Large \bf \system: Blockchain Fault Tolerance}
\author[1,2]{Vincent Gramoli}
\author[3]{Rachid Guerraoui}
\author[1]{Andrei Lebedev}
\author[3]{Gauthier Voron}
\affil[1]{University of Sydney}
\affil[2]{Redbelly Network}
\affil[3]{EPFL}
\date{}

\begin{document}

\maketitle

\begin{abstract}
  Blockchain promises to make online services more fault tolerant due to their inherent distributed nature.
  Their ability to execute arbitrary programs in
  different geo-distributed regions and on diverse operating systems make them an alternative of choice to our dependence on unique software whose recent failure affected 8.5 millions of machines~\cite{CS24}.
  As of today, it remains, however, unclear whether blockchains can truly tolerate failures.

  In this paper, we assess the fault tolerance of blockchain. To this end, we inject failures in controlled deployments of five modern blockchain systems, namely Algorand, Aptos, Avalanche, Redbelly and Solana.
  We introduce a novel \emph{sensitivity} metric, interesting in its own right,
  as the difference between the integrals of two cumulative distribution functions, one obtained in a baseline environment and one obtained in an adversarial environment.
  Our results indicate that (i)~all blockchains except Redbelly are highly impacted by the failure of a small part of their network, (ii)~Avalanche and Redbelly benefit from the redundant information needed for Byzantine fault tolerance while others are hampered by it, and more dramatically (iii)~Avalanche and Solana cannot recover from localised transient failures.
\end{abstract}

\section{Introduction}

One may think that blockchains~\cite{nakamoto_bitcoin_2008} are fault tolerant. They are distributed systems replicated across nodes in geodistributed regions making it unlikely to be affected by a single natural disaster. Their nodes often run different implementations of the same protocol, which reduces the risk of having all nodes experiencing the same bug. In particular, blockchain appears as a promising solution to the recent global CrowdStrike outage~\cite{CS24}. And finally, the owners of these nodes are typically incentivized through cryptoassets to make their node run actively~\cite{RG24}.

Blockchains however are often subject to outages. As an example, Solana experienced 9 outages between September 2021 and February 2023 for a cumulative total of 154.5 hours~\cite{eddie_mitchell_solana_nodate}. In terms of service level agreement (SLA) this translates into offering a service whose availability ($<99\%$) fails to reach two nines, whereas traditional cloud services offer three nines ($\geq 99.9\%$). This questions the ability for blockchain technologies to remain available despite faults.
Unfortunately, previous empirical blockchain comparisons were typically conducted in the absence of failure~\cite{ma_gfbe_2024,gramoli_diablo_2023,noauthor_hyperledger_2024,dinh_blockbench_2017}.

In this paper, we evaluate the fault tolerance of blockchain by
introducing a sensitivity metric and designing  a tool, called \system (\systemlong), that measures the sensitivity of blockchains to dedicated failure patterns.
Finally, we validate \system by
comparing the dependability of five modern blockchain systems:
Algorand~\cite{GHM17}, Aptos~\cite{noauthor_aptos_2022}, Avalanche~\cite{rocket_scalable_2020}, Redbelly~\cite{CNG21} and Solana~\cite{noauthor_solana_nodate}.

\begin{figure}
  \includegraphics[width=\columnwidth, keepaspectratio=true]{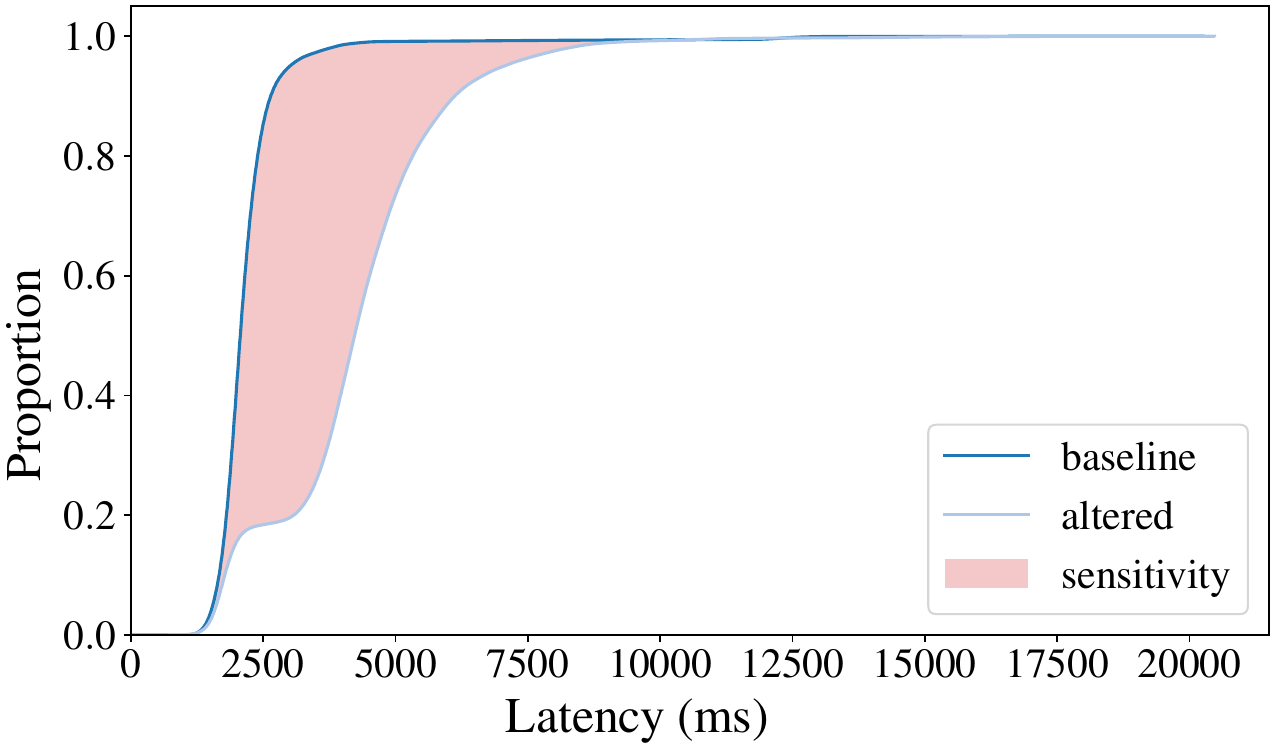}
  \caption{The \emph{sensitivity} of Aptos to failures as the difference in latency distributions between a baseline environment without failure and the altered environment with failures.
  }
  \label{fig:cdf}
  \end{figure}

First, we introduce a new \emph{sensitivity} metric to assess the fault tolerance of a blockchain.
Intuitively, the larger the sensitivity score of a blockchain, the least fault tolerant this blockchain is.
The sensitivity of a blockchain is computed as the difference between the responsiveness of the blockchain in a baseline environment and in an altered environment.
Inspired by the super-cumulative distribution function (SDF) used in economics~\cite{avinash_dixit_stochastic_2007}, which results from the integral of a cumulative distribution function (CDF), we derive the sensitivity as the difference between two SDFs, the ones representing the response times of a blockchain in a baseline environment and in an adversarial environment.

For example, consider \cref{fig:cdf} that depicts two empirical CDFs (eCDFs) of latencies of the Aptos blockchain
(we detail the experimental settings in \cref{sec:sys}).
The first distribution illustrated with the blue curve was observed empirically in a baseline environment without failures. The other distribution illustrated with the light blue curve was observed in an altered environment with failures.
The sensitivity score represented as the light red area is the difference of the areas under the two CDFs, also called SDF.

Second, equipped with this sensitivity metric, we compare for the first time the fault tolerance of different blockchains.
This comparison requires considering blockchains as blackboxes.
To this end, we selected five modern blockchain systems, Algorand, Aptos, Avalanche, Redbelly and Solana, for their ability to tolerate
arbitrary (i.e., Byzantine) failures~\cite{LSP82}.
Given that consensus cannot be solved in the presence of at least $n/3$ permanent failures in an open network where the bound on message delays is unknown~\cite{Gra22}, we study the following properties, where $t<n/3$, for each blockchain:
\begin{itemize}
\item Resilience: the insensitivity to $f=t$ definitive crash (or fail-stop) failures; \item Recoverability: the insensitivity to $f>t$ transient (or crash-recovery) failures;
\item Partition tolerance: the insensitivity to a the partition of $f>t$ nodes; and  \item Byzantine fault tolerance: the insensitivity to $f=t$ arbitrary (or Byzantine) failures.
\end{itemize}

Our results demonstrate that fault tolerance varies greatly with the choice of blockchain system. First, we confirm that all of these blockchains, except Redbelly, are significantly affected by failures.
Second, we show that Avalanche and Solana cannot tolerate transient failures and stop working.
Finally, we show how sending duplicated transactions to cope with Byzantine faults can reduce or improve the responsiveness of blockchain systems.

The paper is organised as follows.
\cref{sec:prelim} presents the background and related work.
\cref{sec:sys} presents our solution and details the experimental settings.
Sections~\ref{sec:resilience}, \ref{sec:recoverability}, \ref{sec:ptolerance} and \ref{sec:btolerance} present respectively the resilience, the recoverability, the partition tolerance and the  Byzantine fault tolerance of the blockchains.
\cref{sec:discussion} discusses our results.
Finally, \cref{sec:conclusion} concludes the paper.

\section{Background and Related Work}\label{sec:prelim}

In this section, we present the previous work. We first introduce the related work and then present each blockchain that we evaluate.

\subsection{Related Work}

The impairments and remedies of dependability of software systems have been studied for more than four decades~\cite{And81,Lap84}. In particular, various books discuss reliability of distributed systems as programming abstraction~\cite{CGR11} or to provide high assurance to applications~\cite{Bir12}. A long series of work studied in particular the \emph{Byzantine} fault tolerance~\cite{LSP82} as the tolerance to arbitrary failures.
It is more recently that blockchain security flaws~\cite{vector76_attack_2021,Fin11} were identified and that the research community started studying blockchain dependability~\cite{NG17}. A long series of blockchain security vulnerabilities can now be found in surveys and books~\cite{CV17,NYG19,Gra22}.

Interestingly, two recent works~\cite{ACS24,KSS24} 
observed vulnerabilities in 
the only two blockchains, Avalanche and Solana, that failed during our experiments. 
First, a theoretical analysis of Avalanche consensus protocols, Snowball and Snowflake, indicate that they do not offer a ``decent'' trade-off between security and performance~\cite{ACS24}.
Second, previous experiments showed that Solana could fork permanently~\cite{KSS24}, however, our observation is different as we noticed that all the nodes of Solana crash after an injection of transient communication delays.

Unfortunately, as of today there is no tool
that allows to systematically compare the fault tolerance of blockchain systems.
Most blockchain evaluation frameworks are focused on performance in fault-free executions~\cite{ma_gfbe_2024,gramoli_diablo_2023,noauthor_hyperledger_2024,dinh_blockbench_2017,nasrulin_gromit_2022}.
They usually measure the latency and throughput of blockchain systems in ideal executions but do not automate the injection of faults to study the sensitivity of these blockchain systems.
Although recent blockchain results were found after injecting crash faults~\cite{VG20} or Byzantine faults~\cite{RG24} in blockchain executions, these injections are typically tailored for a specific blockchain design.
Some results focus exclusively on Byzantine consensus~\cite{AAA24} and ignore other components of a blockchain system.
Other results consider fuzzing~\cite{YTB21,ma_loki_2023,winter_randomized_2023} but require the blockchain code to be analyzed and instumented, making the approach hard to maintain.
As these results do not consider blockchains as blackboxes they cannot be used to compare the fault tolerance of different blockchains on the same ground.

\subsection{Algorand}\label{sec:algorand}

Algorand~\cite{GHM17} is a blockchain that leverages cryptographic sortition through Verifiable Random Functions (VRFs) to randomly select participants for specific roles in the consensus execution. Each participant independently computes a pseudo-random value and a proof, determining their selection for roles such as consensus participant. The Byzantine Agreement ($BA\star$) protocol then uses the consensus participants to propose and validate new blocks, reaching consensus even in the presence of Byzantine faults. This dynamic selection process ensures unpredictable and ever-changing committee membership, enhancing the blockchain security.

To optimize network performance, Algorand adjusts the consensus protocol's timing based on real-time network conditions using Dynamic Round Time~\cite{ciotti_algorands_2024}, ensuring efficient block production while accommodating slower nodes. \emph{Relay} nodes and \emph{participation} nodes have distinct roles, with relay nodes handling data propagation and participation nodes focusing on transaction validation and consensus. However, a single node can fulfill both functions. Transaction propagation is managed through push and pull gossip methods, with push gossip actively broadcasting transactions while pull gossip enabling nodes to request missing transactions, ensuring efficient data synchronization across the network.

\subsection{Avalanche}\label{ssec:avalanche}
Avalanche is a blockchain that builds upon the Snow binary consensus protocol family~\cite{rocket_scalable_2020}. The Snowflake protocol specifically uses three parameters: $k$, $\alpha$ > $k/2$, and $\beta$. Initially, each processor starts with a color, either red or blue. The protocol proceeds in rounds, where in each round, a processor $p$ randomly selects $k$ other processors from the entire population and queries them about their current color. If at least $\alpha$ of the responses differ from $p$'s current color, $p$ switches to that opposite color. If $p$ observes $\beta$ consecutive rounds where at least $\alpha$ of the responses are red (resp. blue), then $p$ decides on red as the final color (resp. blue). With the default parameter values, Avalanche requires at least 80\% of stake to be online for consensus to operate.

Avalanche offers throttling to limit its node resource usage.
Message rate-limiting and connection rate-limiting~\cite{noauthor_avalanchego_nodate} limits the amount of CPU, disk, bandwidth, and message handling a node consumes. 
In particular, the message rate-limiting can be configured based on CPU usage, disk reads/writes, bandwidth usage, and the size and number of unprocessed messages between validators and non-validators, the maximum burst size for bandwidth, and limits on the number of unprocessed messages.
Finally, the connection rate-limiting controls the rate of inbound and outbound peer connections, including the maximum number of connections accepted per second and the frequency of connection attempts. We will discuss how throttling impacts recovery in \cref{sec:throttling}.

\subsection{Aptos}
Aptos~\cite{noauthor_aptos_2022} is a blockchain that builds upon a variant of the HotStuff consensus algorithm~\cite{YMR19} called DiemBFT~\cite{the_diem_team_diembft_2021}, then renamed AptosBFT. In particular, DiemBFT features a view-change mechanism with a quadratic communication complexity instead of the linear approach used in HotStuff, and inherits the cubic communication complexity of the \emph{Practical Byzantine Fault Tolerant (PBFT)} consensus protocol~\cite{castro_practical_1999} that is reached when a faulty is leader or the network is instable. It is thus a \emph{leader-based} blockchain that tolerates up to a third of malicious participants in a partially synchronous environment and that requires view-changes in order to cope with faulty leaders.

Interestingly, Aptos also features the \emph{Block-STM}~\cite{gelashvili_block-stm_2023} design that optimizes the execution of blockchain transactions through Software Transactional Memory (STM), hence the name. In Block-STM, parallel execution leverages multiple threads to execute different transactions concurrently, provided they access distinct memory locations. Aptos execute transactions speculatively to dynamically manage conflicts based on a pre-determined order, but without pre-computing dependencies. When conflicts arise, transactions are aborted and re-executed with their write-sets used to predict and minimize future conflicts.

\subsection{Redbelly}
Redbelly Blockchain~\cite{CNG21} is a scalable blockchain that builds upon the Democratic Byzantine Fault Tolerant (DBFT) consensus algorithm~\cite{CGLR18} that is \emph{leaderless (non leader-based)} and deterministic, and works in a partially synchronous environment.
DBFT has been formally verified with parameterised model checking~\cite{BGL22}, showing that it solves the consensus problem in all possible executions and for any system size.
To enhance scalability further, Redbelly uses a collaborative approach, hence appending a superblock comprising as many valid proposed blocks as possible. This way the number of transactions per appended block can grow linearly with the number of nodes~\cite{CNG21}.

We used the latest version of Redbelly that features the Scalable version of the Ethereum Virtual Machine (SEVM) that runs \emph{decentralised applications (dApps)}
written in Solidity~\cite{THG23} that samples periodically a set of consensus participants among all participants~\cite{Gra22}.
This version was shown to perform well under realistic dApps particularly in a large geo-distributed environment when compared to other modern blockchains~\cite{THG23}. For the sake of security and Byzantine fault tolerance, Redbelly features a library tolerating $f<n/3$ Byzantine failures, called \texttt{credence.js}, for a read operation to return values that are replicated at at least $f+1$ nodes.

\subsection{Solana}\label{ssec:solana}

Solana~\cite{noauthor_solana_nodate} is a blockchain that operates on a pre-determined leader schedule, assigning each validator a specific time \emph{slot}, to produce a block within a larger time frame called an \emph{epoch}. The leader schedule, computed in advance using a pseudo-random algorithm based on data from two epochs prior, ensures validators are chosen proportionally to their stake. This schedule is updated at the end of each epoch and communicated to validators beforehand. A core structure in Solana runtime is the \emph{bank}, which represents the blockchain state at a specific slot, managing transactions, account states, and ensuring adherence to rules during transaction processing. Each bank processes transactions for its assigned slot and, upon completion, finalizes a frozen state that includes a cryptographic hash crucial for network consensus.

Solana runtime includes a mechanism for calculating the \emph{Epoch Accounts Hash (EAH)}, a hash of all accounts, to ensure consistency across validators during each epoch. The EAH is computed between the start and stop slots, typically from one-quarter to three-quarters into an epoch, and integrated into the bank's hash for consensus verification. Notably, Solana does not use a memory pool (or \emph{mempool} for short), forwarding transactions directly to the current and upcoming leaders based on the known leader schedule~\cite{noauthor_retrying_nodate}. If a leader cannot process a transaction in its assigned slot, it passes the responsibility to the next leader.

\section{Measuring Blockchain Sensitivity}\label{sec:sys}

\begin{figure*}
  \includegraphics[keepaspectratio=true, width=\linewidth]{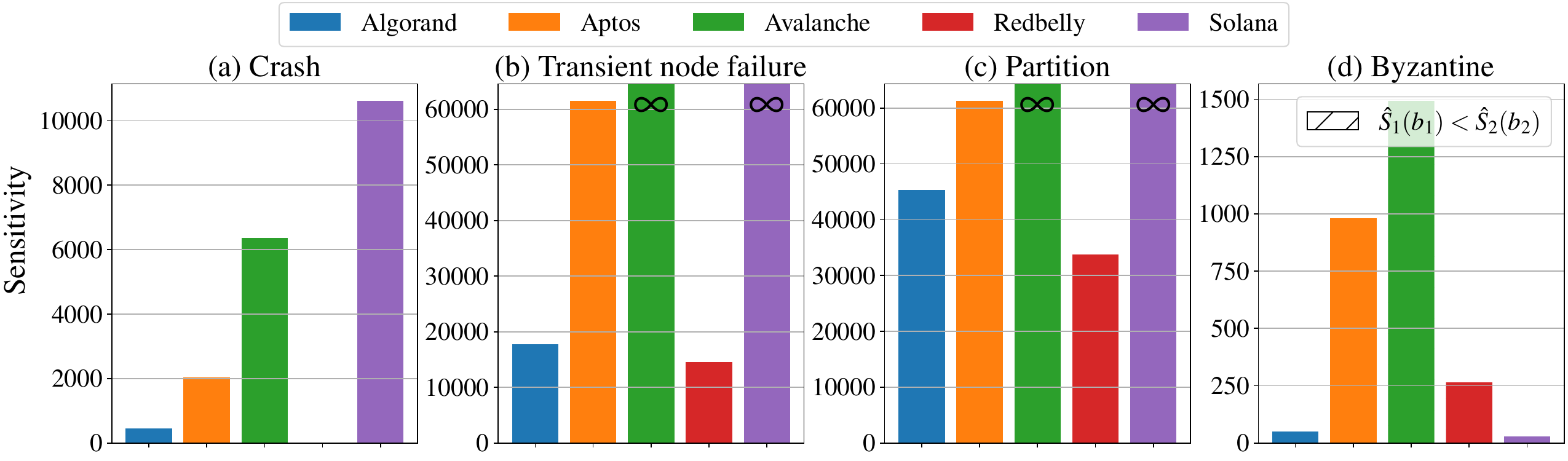}
  {\phantomsubcaption\label{fig:bars-crash}%
   \phantomsubcaption\label{fig:bars-transient}
   \phantomsubcaption\label{fig:bars-partition}
   \phantomsubcaption\label{fig:bars-byz}}%
  \caption{\Score of 5 blockchains
  with $f=t$ crashes, $f=t+1$ transient node failures, transient network partition isolating $f=t+1$ nodes and redundant requests to cope with Byzantine fault tolerance.}
  \label{fig:bars}
  \end{figure*}

\begin{table}
  \caption{Terminology and notation used in the paper.}
  \label{tab:terms}
  \begin{tabularx}{\linewidth}{rX}
    \toprule
    Term & Description \\
    \midrule
    Crash & node is halted and not restarted during the experiment \\
    Transient failure & node is halted and restarted later during the experiment with the same identity \\
    Partition & missing network connectivity between subsets of nodes \\
    Leader & node responsible for proposing a block in the current consensus round \\
    Sensitivity & a measure quantifying the change in transaction latencies in response to variations in the execution environment\\
    Resilience & a measure quantifying the system latency under failures \\
    Recoverability & ability to recover after a transient failure \\
    $f$ & number of failures in an experiment \\
    $t$ & maximum number of failures tolerated by a blockchain \\
    $n$ & number of nodes in a blockchain network \\
    \bottomrule
  \end{tabularx}
\end{table}

In this section, we introduce the \score
to measure the fault tolerance of blockchain systems and explain how we developed a tool called \system to measure it.
We summarize the key terms in \cref{tab:terms}.

\paragraph{Sensitivity score.}
Previous works~\cite{jelasity_bft-bench_2016} rely on three metrics for their evaluation: the latency, the throughput and the downtime.
On the one hand, the latency and throughput metrics quantify the magnitude of the impact of failures on a system and are therefore well suited for permanent failures of a portion of the distributed system.
On the other hand, the downtime quantifies the duration of the effect of failures on the system and is better suited to transient failures.
In order to compare the impact of different types of failures on the same blockchain, \system uses the \emph{\score}, a metric that quantifies both the amplitude and the duration of an effect over a blockchain execution.
We define the \score of a blockchain under a constant workload and in the face of some failures as a function of two latency distributions: one distribution measured in the absence of failures and one with failures.
This function is a mapping from these two distributions to a number defined as the area between the \emph{empirical cumulative distribution function (eCDF)} of the two distributions.

Let \(X\) be a random variable, which can take any value between \(a\) and \(b\), with a cumulative distribution function (CDF) \(F\). The super-cumulative distribution function, or simply super-cumulative~\cite{avinash_dixit_stochastic_2007}, is defined as:

\[
  S(x) = \int_{a}^{x} F(t)dt.
\]

In the setting of transaction latencies, let \((X_1, ..., X_m)\) be the values of a random variable with an eCDF \(\hat{F}(x)=\frac{1}{m}\sum_{i=1}^{m}\mathbf{1}_{X_i \leq x}\), where the sum denotes the number of elements in the sample which are less than or equal to $x$. Then, we adapt the super-cumulative for the eCDF as:

\[
  \hat{S}(x) = \sum_{i=a}^{x} \hat{F}(i).
\]

Consider \(X_1\) as the baseline latency measurements with values between \(a_1\) and \(b_1\), and \(X_2\) as the latency measurements in the altered setting with values between \(a_2\) and \(b_2\), and their corresponding empirical super-cumulatives \(\hat{S}_1\) and \(\hat{S}_2\). The difference \(\hat{S}_1(b_1) - \hat{S}_2(b_2)\) measures the change in the distribution of latencies from the baseline to the altered environment, as seen in \cref{fig:cdf}. However, it is possible that the altered condition improves the performance of a blockchain and decreases transaction latencies~\cite{bronson_metastable_2021, huang_metastable_2022}, in which case \(\hat{S}_2(b_2)\) will be greater than \(\hat{S}_1(b_1)\), producing a negative value of the difference. Since we are measuring sensitivity as the deviation from the baseline, we take an absolute value of the difference, so that the score is always positive, hence the sensitivity score is calculated as \(|\hat{S}_1(b_1) - \hat{S}_2(b_2)|\).

The \score has the following valuable properties that make it an illustrative metric,
as shown in~\cref{fig:cdf}:
\begin{itemize}
\item It measures both the amplitude and the duration of failures effects. Both factors skew the latency distribution of an experiment, resulting in an increased difference between the areas of two empirical super-cumulatives.
\item It is resilient to outliers. Smaller fraction of particular latency values does not contribute significantly to the difference between the areas of two empirical super-cumulatives.
\item It does not require a parameter for interpretation. For example, we do not need a sliding window to explain the score, which may be required to calculate throughput in transactions per second, because the block times might be greater than one second.
\item It is an absolute metric. It allows direct comparison of scores between blockchains and experiments, since it is a function of transaction latencies.
\end{itemize}
Finally, notice that a blockchain that stops committing transactions after a failure event has an infinite \score, which indicates a liveness issue.

\paragraph{\system.}
To calculate \score, we developed \system, a benchmark suite to evaluate blockchains behavior in the presence of faulty processes. \system is built on top of \diablo~\cite{gramoli_diablo_2023}, an open source software to assess the performance of blockchains under realistic but benign workloads.
\system automatically evaluates and compares the ability of several blockchains to tolerate various types of faults.
Specifically, \system evaluates the behavior of blockchains in the face of both permanent and transient failures.
In order to accurately evaluate distributed systems, \diablo is itself a distributed system with two types of machines: the \emph{primary} machine which acts as a central coordinator for the run and many \emph{secondary} machines which simulate clients by submitting transactions to the blockchain processes and waiting for their response.

\paragraph{Observer nodes.}
\system extends the architecture of \diablo in order to control failures during the execution.
Unlike simulated clients, failure events take place on or between the blockchain machines.
Therefore, neither the primary nor the secondary machines are suitable to trigger failures.
Instead, \system uses \emph{observer} processes which run on every blockchain machine and listen to a signal coming from the primary machine.
When the primary decides to trigger a failure on one or many blockchain machines, it broadcasts a signal to the relevant observers.
To implement a crash faults, observer processes simply kill the blockchain process running on their node.
To implement a partition, observer processes use the \texttt{netfilter} interface of their node to drop any IP packet coming from and going to other partitions.
Additionally, observer nodes can end the network partition by removing the \texttt{netfilter} rules or reboot the blockchain process.

\paragraph{Dependability attributes.}
Achieving good performance in the presence of faults, or the \emph{\resilience} metric has received some attention for Byzantine Fault Tolerant state machine replication systems~\cite{clement_making_2009, veronese_spin_2009, aublin_rbft_2013, golan_gueta_sbft_2019}. The metric captures how the system performs with crashed nodes being present in the network, when up to $f$ servers are non-responsive, compared to baseline execution, when all the servers behave correctly.

We measure \emph{\recoverability} of a blockchain as its ability to recover after a transient failure, where the number of failures is greater than the threshold, $f > t$.

From the user perspective, \emph{partitions} display the same behavior as transient faults, as they both result in nodes not being able to exchange the messages. However there is a difference from the implementation perspective. While recovery from transient faults is \emph{active}, since the restarted nodes immediately report their status to the rest of the network after being started, partition recovery can be called \emph{passive}, because the nodes cannot detect that the network connectivity was restored without constant polling.

It is well-known that one client cannot trust the response coming from a single blockchain node: if this blockchain node is Byzantine then the response can be inconsistent~\cite{ruj_ten_2024}.
To cope with this problem in the blockchain systems we studied, where by assumption at most $t$ nodes are Byzantine, the client has to make sure that the same response comes from at least $t+1$ blockchain nodes. This ensures that at least one correct node provided this response.
We therefore study the sensitivity of blockchains to a secure client implementation that compares $t+1$ responses.

\paragraph{Assessing fault tolerance.}
As well as simulating failures, \system differs from \diablo and previous evaluation platforms by implementing \btolerance.
Indeed, a common practice in blockchain client applications is to reach for a single blockchain node and trust it for transmitting the client transactions and relaying the responses from the network.
For example 4 out of the 5 evaluated blockchains (with the exception of Redbelly as we will explain in \cref{ssec:rb-speedup}) provide an SDK for client applications that connect to and trust a single blockchain node~\cite{noauthor_go-algorand-sdkclientv2algodrawtransactiongo_nodate,noauthor_go-ethereumethclientethclientgo_nodate,noauthor_rpcclient_nodate}.
Trusting one specific node effectively brings the number of tolerated Byzantine faults to zero and can lead to devastating cyberattacks~\cite{karame_2021, vector76_attack_2021}.

A common solution is to send the same requests to many, randomly picked, blockchain nodes and compare their responses to detect any faulty response.
Thanks to the deduplication mechanisms, legitimate transactions are executed only once while their results can be observed many times.
This technique however puts an additional load on blockchain nodes as they must deduplicate redundant transactions.
Moreover, this technique likely increases each transaction latency since clients must wait for the slowest of many blockchain nodes instead of one.
We show in \cref{sec:btolerance} that the effect of \btolerance on transaction latency is twofold: it may benefit the transaction latency in mempool-based blockchains, and it may cause redundant transaction execution, even with transaction deduplication mechanisms.

\paragraph{Experimental settings.}\label{sec:expe}

We deployed \system on a distributed system of 15 nodes.
The setup consists of
5 client nodes and 10 blockchain nodes, each client sending native transfer transactions to one blockchain node. Failures are injected on the 5 remaining blockchain nodes that do not receive transactions from clients, this way faulty nodes never receive transactions that they would otherwise lose.
We fixed the total sending rate to 200\,TPS to make sure no blockchains would drop transactions in baseline environments.
In particular, Avalanche capacity is limited to about 357\,TPS because its blocks are produced every 2 seconds and contain a maximum of 714 transactions (as its block limit is 15M gas while the transfer fee is 21K gas).
Each node runs as a virtual machine (VM) of 4 vCPUs and 8\,GB of memory and each of the 5 client nodes sends at the same rate of 40\,TPS to only one of the blockchain node.

To assess the Byzantine fault tolerance of blockchains in \cref{sec:btolerance} we connected each client to 4 blockchain nodes such that each of 5 blockchain nodes has two clients connected to it.
This duplication of requests increased the CPU consumed by the speculative execution of Aptos, which required us to allocated more resources. We thus used VMs with 8 vCPUs and 16\,GB of memory in the Byzantine fault tolerant experiment of each blockchain (\cref{sec:btolerance}).
All the VMs are run on a Proxmox cluster of physical servers, each equipped with 4x AMD Opteron 6378 16-core CPUs running at 2.40 GHz, 256 GB of RAM, and 10 GbE NICs.
We used the following versions of the blockchains: Algorand v3.22.0, Aptos v1.9.3, Avalanche C-Chain v1.10.18-rc.2, Redbelly v0.36.2 and Solana v1.18.1.

In the following sections, we use the sensitivity to different types of failures to evaluate the \Resilience, \Recoverability, \Ptolerance and \Btolerance of blockchain whose results are summarized in \cref{fig:bars}.

\section{Resilience}
\label{sec:resilience}

\begin{figure*}
  \includegraphics[keepaspectratio=true, width=\linewidth]{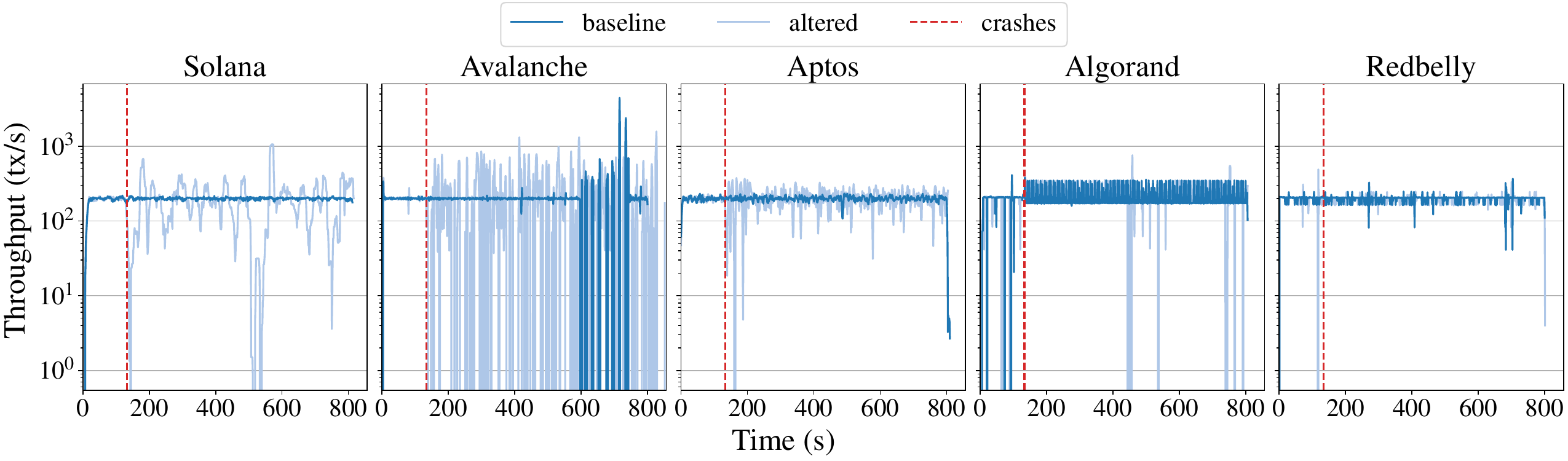}
  \caption{Throughput
  of the 5 blockchains over time as we crash simultaneously $f = t$ nodes at time 133 as indicated by the red dashed line.}
  \label{fig:resilience-tps}
  \end{figure*}

In this section we evaluate the resilience of the 5 tested blockchains.
Our conclusion from the \score to permanent failures is that all blockchains but Redbelly lack resilience. This is due to these blockchains relying on a set of specific servers to make progress at each decision.
In particular, Avalanche and Solana are the least resilient with Solana experiencing higher sensitivity due to better performance in the baseline condition.

\subsection{Assessing resilience}
The test consists of comparing transaction latencies with constant workload in two experiments. The first experiment captures transaction latencies in a fault-free case. The second experiment is divided into a \emph{nominal phase} (preceding any failure) followed by a \emph{crash phase} starting at 133 second timepoint, where we crash $f$ blockchain servers.

\cref{fig:bars-crash} compares the sensitivity of blockchains when $f = t$ nodes experience permanent crash. \cref{fig:resilience-tps} compares the throughput over time in the baseline and altered conditions and offers complementary data to explain the cause of the sensitivity differences between  blockchains.

\subsection{Solana leader impacts performance}
The throughput instability in Solana can be explained by the design decision of not having a mempool~\cite{noauthor_retrying_nodate}. Instead of every node maintaining a temporary storage for transactions, nodes send them directly to scheduled leaders. While such design decision may improve the performance in the best case scenario, when the scheduled leader processes the transactions, it leads to a snowball effect when a scheduled leader is non-responsive. With the constant workload, the new leader has to process a higher volume of transactions since one or more scheduled leaders are down. Hence, with crashed nodes being present in the leader schedule, we observe periods of low throughout when the scheduled leader is down, and throughput peaks when the transactions are processed by a responsive node, resulting in higher latencies, and therefore higher score.

\subsection{Avalanche throttling leads to instability}\label{sec:throttling}

In \cref{fig:resilience-tps}, we observe that Avalanche throughput is unstable. This is explained by its throttling mechanism.
Several voting rounds should successfully pass in succession to commit a block. Since nodes are sampled for every voting round from all the nodes in the network, in the presence of crashes, faulty nodes may be included in the samples as well.

 Intuitively, even with node crashes, the repeated sampling should allow the network to come to an agreement on a block. However, as we mentioned in \cref{ssec:avalanche} the current implementation includes multiple layers of message throttling based on CPU usage, bandwidth, and number of messages. The nodes exchange the messages, including transactions and consensus data, and the messages are first stored in queues before being processed.
With the 200 TPS constant workload and default throttling settings, the nodes do not process the messages, even though the messages are sent and received over the network. The nodes consuming their respective CPU quotas cause the messages not to be processed, leading to messages not reaching the consensus module and increasing the throughput instability.

Additionally, we discovered a previously reported bug~\cite{noauthor_pull_nodate} with the help of \system. However, after running the experiments with a fixed version, we did not observe a measurable performance improvement because throttling has greater impact on the transaction latency and throughput.

\subsection{Aptos mitigates the leader impact}
Aptos displays significant oscillations immediately after the crashes, however in contrast to Solana and Avalanche, the throughput instability reduces in about 82 seconds at 215 second timepoint. This behavior matches the description of the DiemBFT protocol~\cite{the_diem_team_diembft_2021}. While we tested a network of 10 nodes and observed noticeable performance drop, we can expect the performance to get increasingly worse with the growth of the network size.

\subsection{Algorand adapts slowly to sudden failures}
Algorand throughput depends on the timing parameters, which are calculated dynamically based on the observed round finalization times. Since the servers are selected using the VRF of $BA^\star$, samples may include crashed nodes, which increase the round finalization time. Initially, default timing parameters are used, which are then reduced, explaining the throughput increase after approximately 133 seconds have passed since the start of the experiment. In the presence of crashes, there are periods when the decreased timing parameters are reset to their default values, which reduces the average throughput and increases transaction latency. Such periodic increases in latency are reflected in the score.

\subsection{Redbelly eradicates the leader impact}
Redbelly is not affected by the presence of $f = t$ crashes. The reason is that Redbelly uses the leaderless consensus algorithm, called DBFT.
In particular, Redbelly features a Byzantine fault tolerant binary consensus algorithm and a classic reduction from the multi-value consensus problem to the binary consensus problem.

First, Redbelly is not affected by the slow reponsive node that affects Solana because no individual slow node can significantly slow down the DBFT consensus protocol.
More specifically, even if its binary consensus algorithm uses a weak coordinator to break symmetry,
a faulty weak coordination does not prevent the DBFT
consensus algorithm from converging towards a decision~\cite{CGLR18}.

Second, Redbelly does not show the sign of oscillation of Aptos. As confirmed by previous results~\cite{VG20} this is due to DBFT being less
impacted than HotStuff-like protocols (including DiemBFT) when their leader crashes.
As a result, the leaderless consensus protocol reduces the effect that of not only a single slow node but also a single crashed node have on the overall blockchain execution.

\section{Recoverability}
\label{sec:recoverability}

In this section, we evaluate the recoverability of the 5 tested blockchains.
We can conclude from our results that two blockchains, Avalanche and Solana, cannot recover from
a number of transient node failures.
The other blockchains can recover from transient node failures but with varying speeds.

\subsection{Assessing recoverability}
To test if a blockchain can recover we inject transient failure and  run an experiment in 3 phases. Similarly to the resilience experiment (cf. \cref{sec:resilience}), we start with the nominal phase with no failures. In the fault phase starting at 133 seconds, we halt $f = t + 1$ nodes for 133 seconds. After the fault phase, we continue with the measure phase with $f = 0$ by restarting the nodes.

\begin{figure*}
\includegraphics[keepaspectratio=true, width=\linewidth]{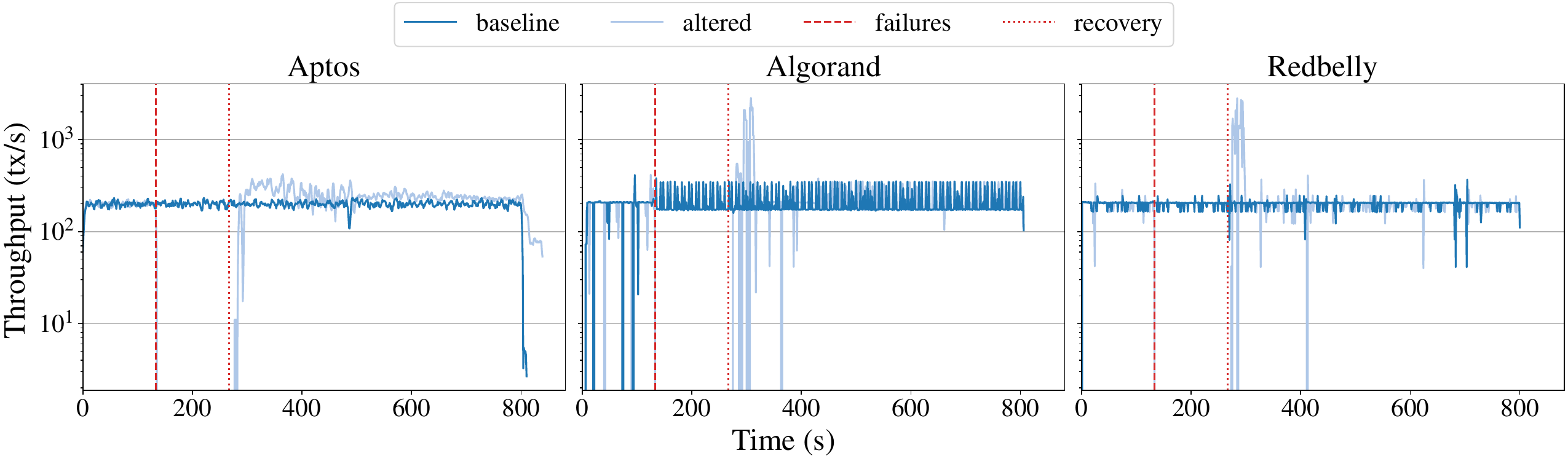}
\caption{Throughput of
 the 5 blockchains over time as we transiently stop $f > t$ nodes at time 133 as indicated by the dashed red line and as we recover them at time 233 as indicated by the dotted red line.
}
\label{fig:recoverability-tps}
\end{figure*}

\cref{fig:bars-transient} shows the score for each blockchain with the presence of a transient failures of $f = t + 1$ nodes, and \cref{fig:recoverability-tps} throughput over time in the baseline and altered conditions. When the blockchain is unable to recover from the crashes and restore liveness, \textbf{$\infty$} symbol is displayed in \cref{fig:bars-transient} instead of the corresponding bar.

\subsection{Solana generalized failure}
\label{ssec:solana-fail}

We observed, somehow surprisingly, that the transient failures of some nodes crash all the nodes of Solana.
Our in-depth investigations led to conclude that this is related to a  bug that prevents a node from synchronizing its state with another.

\sloppy{As explained in \cref{sec:prelim}, Solana requires an Epoch Accounts Hash (EAH) to be calculated for the consensus. The panic observed in the Solana validator node stems from an unmet precondition during the EAH calculation process~\cite{noauthor_solanadiscord_nodate}, specifically within the \texttt{wait\_get\_epoch\_accounts\_hash} function. This function is responsible for ensuring that the EAH calculation is either completed or correctly initiated at the expected point in the epoch.
The panic occurs because
of the ordering of two parallel events: the EAH calculation and the EAH integration.
In particular,
no EAH calculation was started or in-flight when the bank (described in \cref{ssec:solana}) attempted to integrate the EAH into the bank hash at three-quarters (3/4) of the epoch duration, which is a critical part of the consensus process.
}

By investigating the stack trace,
no bank was rooted at the beginning of the epoch, preventing the EAH calculation from starting.
As a result, when the bank reached 3/4 of the epoch duration, it was unable to complete the EAH process that had not started. This led to a critical failure because the bank cannot retroactively initiate the EAH calculation, making it impossible to fulfill the required consensus step.

After identifying the cause of the panic, we went to Solana discord channel and found that Solana needed at least 360 slots per epoch~\cite{noauthor_validator_nodate} while being configured with a smaller amount of slots. The reason is that Solana needs enough time to compute the EAH and to root the relevant bank before the 3/4 mark. Given that rooting can sometimes take up to 150 slots and the freeze-to-rooting process requires at least 32 slots, a buffer is necessary to ensure everything completes correctly. Therefore, the minimum epoch length must be around 360 slots to allow this process to happen without causing a panic. This ensures the EAH computation and rooting are completed in time, preventing errors that could occur if an epoch were too short.

The default epoch duration for the development cluster is 8192 slots. However, with the deployment scripts provided in the Solana repository, genesis block is generated with \texttt{enable-warmup-epochs} flag, which shortens the first 8 epochs to progressively smaller slot counts, beginning with 32 slots in epoch 0. These warm-up epochs follow a geometric progression, where the number of slots doubles with each epoch. The epoch size returns to normal (8192 slots) after the warm-up period. The first full-length epoch occurs after 54m24s, and prior to that, each epoch's duration is much shorter. We introduce transient faults to the system at 133 seconds during one of the warm-up epochs, specifically when the number of slots per epoch is still under 360, leading to the described issue.

\subsection{Avalanche lack of liveness}
\label{ssec:avalanche-fail}

In Avalanche, we did not experience a generalised outage like in Solana, however,
Avalanche's throttling implementation described in \cref{sec:throttling} also prevents the network from reaching consensus.

More specifically,
the \texttt{InboundMsgThrottler} of Avalanche contains multiple throttler structures, among which the  CPU quota-based throttler, \texttt{cpuThrottler}, and the message buffer-based throttler, \texttt{bufferThrottler}, are of particular interest.

\sloppy{First, the \texttt{cpuThrottler} leverages functions to block the CPU consumption of incoming message processing. When a message arrives, the \texttt{systemThrottler.Acquire} function checks if there is enough CPU quota available based on the current CPU usage tracked by \texttt{cpuResourceTracker.Usage}. The decision is also influenced by the \texttt{targeter.TargetUsage}, which sets a CPU usage threshold. If the current usage approaches or exceeds this target, \texttt{cpuThrottler} blocks further processing of messages, effectively throttling them until CPU resources are freed, preventing the system from exceeding the allocated CPU quota.}

\sloppy{Second, the \texttt{bufferThrottler}
rejects messages depending on the buffer availability with \texttt{inboundMsgBufferThrottler.Acquire}. When the system buffers are saturated---typically because the CPU throttling has prevented messages from being processed---the buffer throttler restricts further intake of messages. This backpressure mechanism ensures that incoming messages do not overflow the system when the processing pipeline is already overwhelmed and cannot clear the buffers efficiently.}

We observed from the logs that the messages were successfully sent and received by the nodes during the experiments, but the throttling prevented them from being processed in a timely manner, resulting in no new blocks being agreed upon. Note that Avalanche is known to require a variant of asynchrony with some form of synchrony for liveness~\cite{rocket_scalable_2020}. What this experiment seems to demonstrate is that Avalanche stops working when some messages arrive 2 minutes late.

\subsection{Algorand and Redbelly recovery}\label{sec:alg-rb-recovery}
From \cref{fig:bars-partition} and \cref{fig:bars-transient}  Algorand and Redbelly display the best behaviors among the studied blockchains when the number of failed nodes exceeds the fault tolerance threshold for a short period.
After the crashed nodes are restarted (at 266 seconds), we quickly observe a sharp peak in throughput. This peak corresponds to processing the accumulated backlog of transactions during the downtime. For the rest of the experiment, throughput and latencies match the measurements acquired during the nominal phase (i.e., the first 133 seconds).

\subsection{Aptos unrecoverable performance drop}
Among three remaining blockchains, Aptos is most significantly impacted by the loss of liveness in the presence of $f = t + 1$ transient failures. While the network starts to create and commit new blocks shortly after restarting the crashed nodes, the transaction throughput does not return to the values observed in the nominal phase. Compared to Algorand and Redbelly, the throughput amplitude is significantly lower for Aptos, meaning that it cannot process the pending transactions as fast as Algorand or Redbelly. Furthermore, since we record committed blocks after the end of the experiment, we can observe that the blocks are still being created for a certain period of time. Therefore, we can conclude that Aptos fails to clear the backlog of transactions accumulated during the downtime, and the performance remains degraded for the rest of the experiment, displaying increased transaction latencies.

\section{Partition Tolerance}
\label{sec:ptolerance}

In this section we evaluate the partition tolerance of the 5 blockchains.
We conclude that proactive detection of link failures can improve the blockchain performance.
We also confirm that the blockchains that could not tolerate transient node failures cannot tolerate partition either.

\begin{figure*}
\includegraphics[keepaspectratio=true, width=\linewidth]{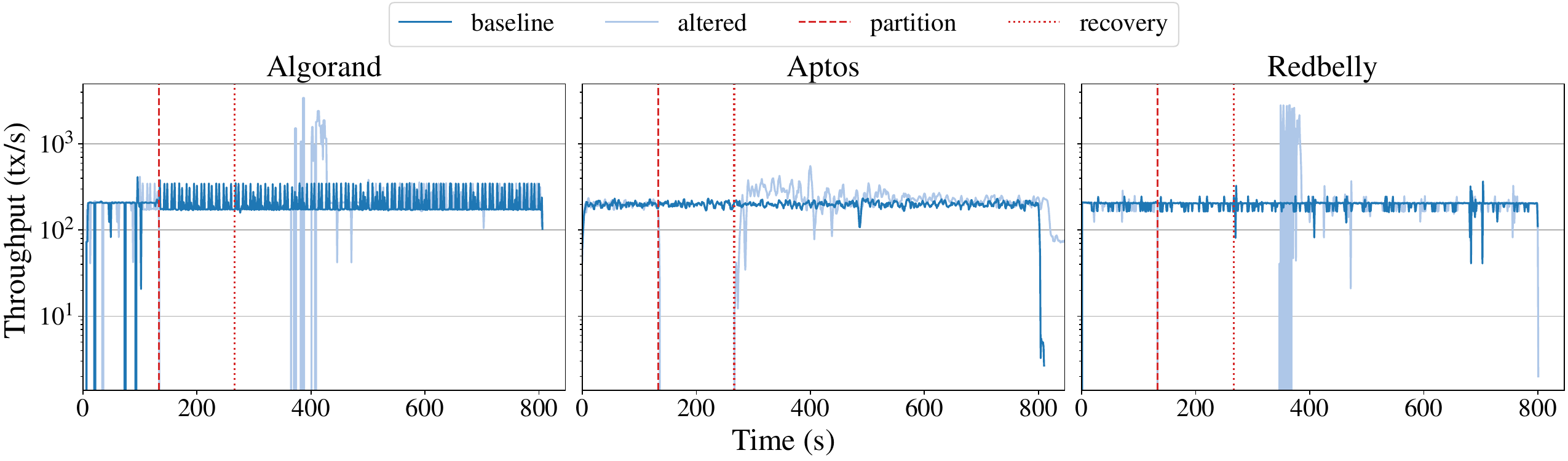}
\caption{Throughput of
the 5 blockchains over time as we transiently partition $f > t$ nodes at time 133 as indicated by the dashed red line and as we stop the partition at time 233 as indicated by the dotted red line.
}
\label{fig:partition-tps}
\end{figure*}

\subsection{Measuring partition tolerance}
To test the potential difference in the performance in the presence of network partitions, we replace the fault phase in the experiment described in \cref{sec:resilience} with the blockchain network being partitioned, with $f = t + 1$ nodes being in the smaller partition.

\sloppy{We used Linux traffic control subsystem facilities to create a transient link failure, or a network partition. First, we used \texttt{tc qdisc add dev eth2 root handle 1: prio} to establish a priority queuing discipline at the root of the \texttt{eth2} interface. Next, we used \texttt{tc filter add} to define a set of filters that match IP packets with a destination of IP addresses of the nodes we wanted to disconnect and direct them to the third priority band (\texttt{flowid 1:3}). Then, we applied \texttt{tc qdisc add} to introduce a \texttt{netem qdisc} on \texttt{flowid 1:3}, simulating 100\% packet loss for the matched traffic. Finally, we used \texttt{tc qdisc del dev eth2 root} to remove all traffic control configurations on \texttt{eth2}.}

We report the sensitivity scores in \cref{fig:bars-partition}, and show the corresponding throughput over time in \cref{fig:partition-tps}.

\subsection{Solana and Avalanche lack of recovery}
Solana and Avalanche fail to recover and restore liveness after the partition, hence we show $\infty$ symbol in addition to the corresponding bar.
The issues that make Solana and Avalanche fail to recover after a partition are the same as with the transient node faults previously described in \cref{ssec:solana-fail,ssec:avalanche-fail}. In Solana, the EAH calculation, which occurs after the network connectivity is restored, causes all the nodes to crash. In Avalanche, the throttling mechanism prevent the nodes from exchanging the transactions and reaching consensus.

As we explained below (\cref{sec:alg-rb-timeouts,sec:aptos-timeout}), the other three blockchains, Algorand, Aptos and Redbelly,
that recover from transient node failures, show different scores under network partition.

\subsection{Algorand and Redbelly timeouts}\label{sec:alg-rb-timeouts}

The score of Algorand and Redbelly observed under network partition is higher than their respective score obtained under transient node failures in \cref{sec:recoverability}.
In particular, if we compare \cref{fig:recoverability-tps} to \cref{fig:partition-tps}, we observe
that the recovery time of Redbelly increased
from 7 to 81 seconds while the recovery time for Algorand increased from 9 to 99 seconds.

After investigating the code of Algorand and Redbelly, we concluded that the recovery time was a function of specific timeouts. After these timeouts expire, the nodes attempt to reconnect with each other.
By looking closer at the code of Redbelly and discussing with its developers, we noticed that an existing \texttt{MaxIdleTime} timeout variable of 30 seconds, could help speedup the recovery further.

\subsection{Aptos backoff time for quick recovery}\label{sec:aptos-timeout}

By contrast with Algorand and Redbelly,
Aptos displays the same sensitivity to nodes being under a transient failure (\cref{fig:bars-transient}) and to a network partition (\cref{fig:bars-partition}). Such a contrast can be explained by different implementation strategies for detecting the network connectivity being restored.
Aptos checks peer connectivity every 5 seconds by default. Validator connections are maintained with exponential backoff waiting time with the base value of 2 seconds. A timeout for the connection to open and complete all of the upgrade steps is 30 seconds. Such parameters allow quick connection recovery after the network partition is restored compared to Algorand and Redbelly.

\section{Byzantine Fault Tolerance}
\label{sec:btolerance}

In this section we measure the sensitivity of blockchain systems to Byzantine fault tolerant requests.
We conclude that with the more secure client, the request redundancy can benefit the transaction latency in mempool-based blockchains. In addition, the transaction deduplication and execution mechanisms should be tested in the process of the development.

\subsection{Assessing Byzantine fault tolerance}\label{ssec:redundant-cllient}

Assessing Byzantine fault tolerance
is difficult because of the infinite number of Byzantine executions. In order to assess Byzantine fault tolerance, we thus implement a secure client that compares the response from $t+1$ blockchain nodes before returning the aggregated answer to the application layer.

To test whether a blockchain performance is impacted by the secure client implementation, we sent the same transaction to 4 different nodes instead of a single node, and reported the transaction as being committed only after all 4 nodes have responded. We used 4 nodes since it is the maximum value for $t+1$ with $n=10$ across the blockchains under test. The experiment has a single phase during which we use the modified client. As discussed in \cref{sec:sys}, we deployed VMs with 8 vCPUs and 16 GB RAM in order to prevent dropped transactions in Aptos, since a redundant client causes extra CPU load on the nodes, as explained later in \cref{ssec:speculative}.

We depict in \cref{fig:bars-byz}, the \score for each blockchain with a client connected to 4 nodes.

\subsection{Algorand and Solana remain unchanged}
The low sensitivities of Algorand and Solana in \cref{fig:bars-byz} indicate that neither Algorand nor Solana are significantly affected by the redundant requests from the client.

In Algorand, since we used a fully-connected network in our experiments, where nodes function as both relay and participation nodes, we do not observe the expected reduction in transaction latency and throughput improvements when a 4-connected redundant client is used. Each node maintains a transaction pool, holding transactions in memory before proposing them in a block. Additionally, push and pull gossip methods propagate these transactions across all connected nodes. However, since every node is directly connected and plays dual roles, the network lacks the hierarchical or segmented structure that typically benefits from such optimizations. Consequently, the benefits of reduced latency and enhanced performance are mitigated by the inherent redundancy and uniform connectivity, leading to minimal impact on overall network efficiency.

In Solana, sending a transaction to multiple nodes does not help reduce latency, increase throughput, or improve performance because of its mempool-less architecture. As discussed in \cref{ssec:solana}, Solana uses an approach where transactions are directly forwarded to the expected leaders based on a pre-determined leader schedule. This process eliminates the need for a mempool, where transactions typically wait to be processed by validators. As a result, broadcasting a transaction to multiple nodes is redundant since all nodes will ultimately route the transaction to the same set of leaders, who will anyway handle it according to the network deterministic leader schedule.

\subsection{Aptos speculative execution drawback}\label{ssec:speculative}
The root cause of the performance degradation in Aptos on \cref{fig:bars-byz} seems to come from the speculative execution of Block-STM transactions~\cite{gelashvili_block-stm_2023}. We observe the following differences in blockchain node logs between the baseline experiment with a single client, and the altered case with the redundant client connected to 4 nodes.
In both executions, first, a transaction is added to the mempool, reported by a log message from \texttt{Mempool::add\_txn} function. Then, a transaction is committed and removed from the mempool, reported by a log message from \texttt{Mempool::log\_commit\_transaction} function. However, in the altered execution, we additionally observe a log message from \texttt{SpeculativeEvent::dispatch} 10 milliseconds later with \texttt{SEQUENCE\_NUMBER\_TOO\_OLD} error, since the transaction is already committed. This means that some transactions are getting processed at least twice with the redundant client, causing additional load to the nodes.

\subsection{Redbelly speedup}\label{ssec:rb-speedup}

As discussed in \cref{sec:sys}, the sensitivity score is always positive as it represents the difference expressed as the absolute value, \(|\hat{S}_1(b_1) - \hat{S}_2(b_2)|\), between the area of the baseline environment $\hat{S}_1(b_1)$ and the area of the altered environment $\hat{S}_2(b_2)$.
Without this absolute value, the sensitivity could be negative, if the altered environment was offering lower latencies than the baseline environment.
This interesting scenario is observed here, because the altered environment benefits Redbelly more than the baseline environement, i.e., \(\hat{S}_1(b_1) < \hat{S}_2(b_2)\), as depicted by the crossed bar in \cref{fig:bars-byz}.

The slight latency drop
that Redbelly experiences in the altered environment is probably due to the superblock optimisation it uses to solve the Set Byzantine Consensus~\cite{THG23}.
In particular, as opposed to  classic blockchains that decide one of the proposed blocks, Redbelly decides a superblock that combines the valid transactions from all the proposed blocks.
As the altered environment sends
the same transactions to multiple nodes, it can increase the chances of a transaction being included in the superblock slightly earlier.

Finally, it is important to note that Redbelly already offers its specific recommended library to ensure Byzantine fault tolerance~\cite{Gra22c}.
This library called \texttt{credence.js} guarantees to a client that the responses it obtains had identical hashes across $t+1$ replicas.
We decided not to use this library and to use our modified client described in \cref{ssec:redundant-cllient} to obtain a fair comparison with other blockchains.

\subsection{Avalanche slower sequential execution}
Avalanche experiences the largest  sensitivity among all the blockchains (\cref{fig:bars-byz}). Interestingly, however, Avalanche, just like Redbelly in \cref{ssec:rb-speedup} benefits from the redundant requests sent by the client to cope with Byzantine fault tolerance, which is indicated in \cref{fig:bars-byz} with \(\hat{S}_1(b_1) < \hat{S}_2(b_2)\). This is because this redundancy compensates the reordering of transactions and the throttling effects as we explained below.

First, the leader rotation combined with the gossip implementation can increase the latency. The reason is that transactions have to be executed in the order of their issuance---this is enforced like in other blockchains by assigning a transaction with a nonce that counts the previous transactions issued by the same account owner.
For a transaction of an account owner to be executed, all its previous transactions (with lower nonces) must first reach the leader. This can take a long time depending on the leader rotation order and the gossip protocol.
Consider a client submitting two transactions with nonces 1 and 2 to the network. The node receiving the transactions may not become a leader for a certain period of time given the leader rotation.
The Avalanche protocol relies on a gossip-based protocol and for every invocation that pushes a message, Avalanche collects a set of transactions from a HashMap data structure. These transactions are collected in a for loop~\cite{noauthor_corethcoretxpoollegacypoollegacypoolgo_nodate} from these HashMaps whose keys do not determine the order~\cite{noauthor_go_nodate}. This is why the transaction with the lowest nonce can be delayed. But, sending a transaction to multiple nodes as it is done for the sake of Byzantine fault tolerance increases the chances of a transaction becoming immediately available for the current leader and being included in a block earlier.

Second, the throttling mentioned in \cref{sec:resilience} prevents the gossip messages from being immediately processed by the nodes. The message queues being handled by throttling include different internal messages, for consensus and transactions. Since clients are processed separately from the blockchain nodes, using a redundant client allows to mitigate the negative performance impact caused by throttling and improve transaction latency.

\section{Discussion}\label{sec:discussion}

In this section, we give a summary of our blockchain dependability results and discuss the limitations of our approach.

\begin{figure}
\includegraphics[width=\columnwidth, keepaspectratio=true]{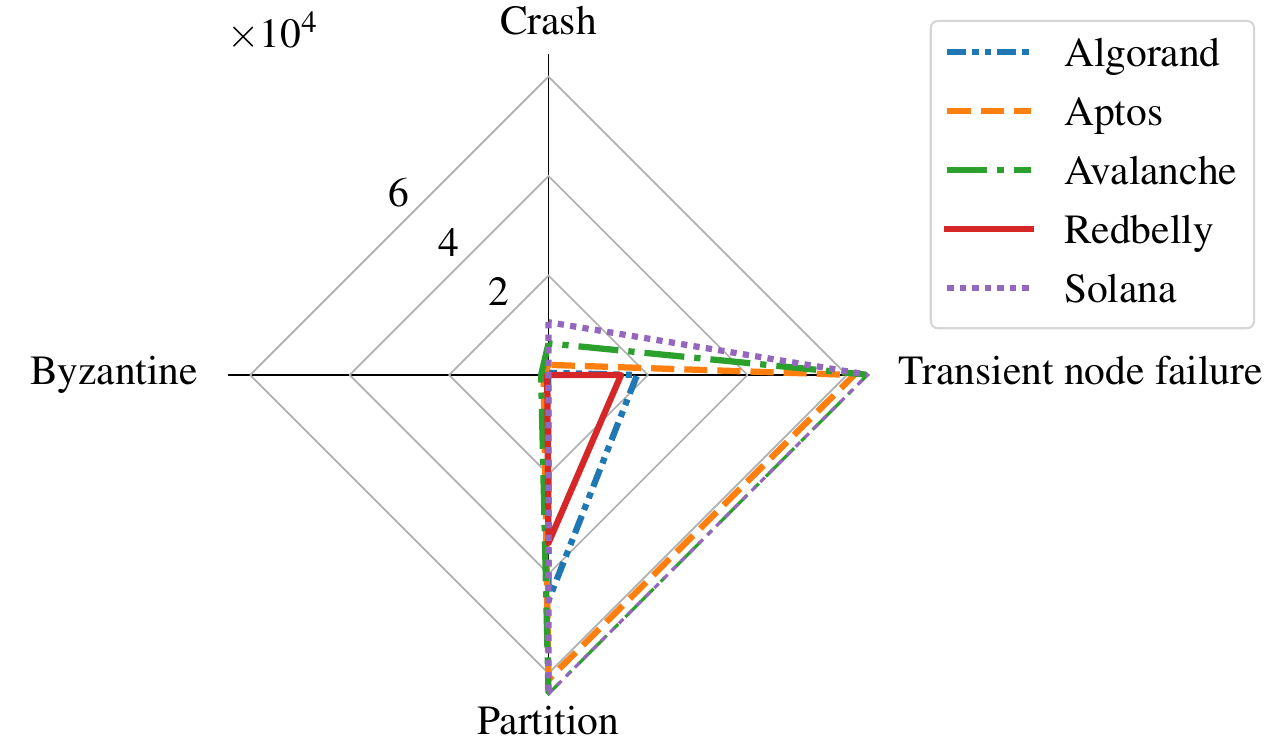}
  \caption{The sensitivity of the tested blockchains to partition, crash, Byzantine and transient failures.}
  \label{fig:radar}
\end{figure}

\paragraph{Synthesizing the results.}
To get a better understanding of the dependability of the 5 studied blockchains, we report all the sensitivity scores we measured in the previous sections on Figure~\ref{fig:radar}.
The scores are displayed in a radar chart with four dimensions representing their sensitivity to crash failures, transient node failures, partitions and Byzantine failures.
For each type of failures, the higher the reported value, the higher the sensitivity to this type of failure.

The first observation is that generally blockchains are more sensitive to transient failures than permanent failures. First, the blockchains are generally very sensitive to transient failures whether these are link failures, as illustrated by partitions, or node failures, as illustrated by a crash followed by a recovery of individual nodes.
Second, they are not as sensitive to the permanent failures. In fact, one can barely see the sensitivity to Byzantine failures and the sensitivity to crash failures is significantly lower than the sensitivity to partitions and transient node failures.
After a specific number of transient failures, some blockchains (Solana and Avalanche) could not even recover.
Note that because of the fault tolerance threshold of these blockchains, we introduced less permanent failures ($f\leq t$) than we have introduced transient failures ($f>t$). We can conclude that Solana and Avalanche are likely tuned to only support as many slowly responsive nodes as they can afford permanent failures.

The second observation is that some blockchains (Algorand, Aptos and Redbelly) recover as one would expect but with varying speeds. In particular, Aptos recovers particularly slowly from transient failures as Algorand and Redbelly recover signficantly faster. Finally Redbelly is the fastest blockchain to recover. This can be due to two things. First, the extensive research done around its dependability under Byzantine attacks~\cite{CNG21} and flooding attacks~\cite{THG23} required it to cope with adversarial network scenarios impacting the delays of messages.
Second, it was shown to have better performance than most of the other blockchains~\cite{THG23} not only due to a reduction in the number of verifications that it needs but also due to its superblock optimization that commits a large batch of accumulated transactions faster.

\paragraph{Limitations of our approach.}
Our work is a first attempt towards evaluating blockchain dependability. We focused on only five blockchains that are claimed to be Byzantine fault tolerant but there are many more blockchain proposals with the same claim that we could evaluate as well, however, we were unsure of their level of maturity. It is relatively easy to add other blockchains to our framework and we encourage the research community to reuse our results and measure the sensitivity of other blockchains.

The settings of our experiments may seem far from being realistic because blockchain networks are generally of larger scales than our 15-node distributed system and the distance between nodes is generally larger than within a cluster. However, recent results showed that one can get a deep understanding of the performance (both in latency and throughput) of a blockchain at large scale even when deployed in a much smaller environment~\cite{LG23}.

Finally, the workload that we use for our experiments only sends native transfer transactions at a constant rate of 200\,TPS, which is not representative of realistic   fluctuating workloads, request bursts or demanding workloads. The reason for using simple transactions is that some blockchains are unable to support complex smart contract invocations because of the amount of gas they would consume~\cite{gramoli_diablo_2023}. The reason for using a relatively low sending rate is that some blockchains would lose transactions if the sending rate is too high~\cite{gramoli_diablo_2023}, which typically incurs congestion bottlenecks. Limiting these undesirable effects allowed us to better observe the impact of failures on latencies, which was crucial to measure sensitivity.

\section{Conclusion}\label{sec:conclusion}

We presented the first fault tolerance comparison of blockchain systems.
To this end, we introduced a new \emph{sensitivity} metric, interesting in its own right, derived from the super-cumulative distribution functions of service response times.
This sensitivity metric allowed us to measure \textit{(i)}~the resilience, \textit{(ii)}~the recoverability, \textit{(iii)}~the partition tolerance, and \textit{(iv)}~the Byzantine fault tolerance of five modern blockchain systems:
Algorand, Aptos, Avalanche, Redbelly and Solana.
Our future work includes evaluating Byzantine fault tolerance using recommended specialized client libraries, such as \texttt{credence.js} for Redbelly.

\section*{Acknowledgments}

We wish to thank the developers of the blockchains who helped us understand the cause of our results: The Avalanche developers confirmed the bug mentioned in \cref{sec:throttling} we also found. The Redbelly developers confirmed that the recovery time could be shorten with the 30-second timeout as discussed in \cref{sec:alg-rb-timeouts}. The head of the developer relations at Algorand helped us understand the throughput variations of \cref{sec:algorand} before and after the dynamic round time.
 This research is supported under Australian Research Council Future Fellowship funding scheme (project
  number 180100496) entitled ``The Red Belly Blockchain: A Scalable Blockchain for Internet of Things''.

\bibliographystyle{plain}
\bibliography{references}

\end{document}